\documentclass[usenatbib]{mn2e}

\usepackage{epsfig}
\usepackage{mathrsfs,aas_macros}

\def\Lya{Ly$\alpha$~}

\def\HI{\hbox{H$\,\rm \scriptstyle I\ $}}
\def\HII{\hbox{H$\,\rm \scriptstyle II\ $}} 
\def\HeI{\hbox{He$\,\rm \scriptstyle I\ $}}
\def\HeII{\hbox{He$\,\rm \scriptstyle II\ $}}
\def\HeIII{\hbox{He$\,\rm \scriptstyle III\ $}}

\title[The IGM temperature around a quasar at $z=6$]{A first direct measurement
  of the intergalactic medium temperature around a quasar at
  $z=6$}

\author[J.S. Bolton et al.] {James S. Bolton$^{1}$, George
  D. Becker$^{2}$, J. Stuart B. Wyithe$^{1}$, Martin
  G. Haehnelt$^{2}$ \newauthor \& Wallace L.W. Sargent$^{3}$\\ $^1$ School of Physics, University of
  Melbourne, Parkville, VIC 3010, Australia \\ $^2$ Kavli Institute
  for Cosmology and Institute of Astronomy, Madingley Road, Cambridge,
  CB3 0HA \\ $^3$ Palomar Observatory, California Institute of Technology, Pasadena, CA 91125, USA}

\begin{document}

\date{}

\maketitle

\label{firstpage}

\begin{abstract}

The thermal state of the intergalactic medium (IGM) provides an
indirect probe of both the \HI and \HeII reionisation epochs.  Current
constraints on the IGM temperature from the \Lya forest are restricted
to the redshift range $2 \leq z \leq 4.5$, limiting the ability to
probe the thermal memory of \HI reionisation toward higher redshift.
In this work, we present the first direct measurement of the IGM
temperature around a $z=6$ quasar by analysing the Doppler widths of
\Lya absorption lines in the proximity zone of SDSS~J0818$+$1722.  We
use a high resolution ($R= 40\,000$) Keck/HIRES spectrum in
combination with detailed numerical modelling to obtain the
temperature at mean density, $T_{0}=23\,600\pm^{5000}_{6900}\rm\,K$
($\pm^{9200}_{9300}\rm\,K$) at 68 (95) per cent confidence assuming a
prior probability $13\,500\rm\, K \leq T_{0} \leq 38\,500\rm\,K$
following \HI and \HeII reionisation.  This enables us to place an
upper limit on the redshift of \HI reionisation, $z_{\rm H}$, within
$33$ comoving Mpc of SDSS~J0818$+$1722.  If the quasar reionises the
\HeII in its vicinity, then in the limit of instantaneous reionisation
we infer $z_{\rm H}<9.0$ $(11.0)$ at 68 (95) per cent confidence
assuming photoheating is the dominant heat source and that \HI
reionisation is driven by ionising sources with soft spectra, typical
of population II stars.  If the \HI and \HeII in the IGM around
SDSS~J0818$+$1722 are instead reionised simultaneously by a population
of massive metal-free stars, characterised by very hard ionising
spectra, we obtain a tighter upper limit of $z_{\rm H}<8.4$ $(9.4)$.
Initiating reionisation at higher redshifts produces temperatures
which are too low with respect to our constraint unless the \HI
ionising sources or the quasar itself have spectra significantly
harder than typically assumed.

\end{abstract}
 

%%%%%%%%%%%%%%%%%%%%%%%%%%%%%%%%%%%%%%%%%%%%%%%%%%%%%%%%%%%%%%%%%%%%%	
%%%%%%%%%%%%%%%%%%%%%%%%%% SECTION 1 %%%%%%%%%%%%%%%%%%%%%%%%%%%%%%%%
%%%%%%%%%%%%%%%%%%%%%%%%%%%%%%%%%%%%%%%%%%%%%%%%%%%%%%%%%%%%%%%%%%%%%

\section{Introduction}

The discovery of $z \simeq 6$ quasars within the last decade has led
to several important advances in our understanding of the high
redshift Universe.  One area where the impact of this work has been
especially significant is the study of the high redshift intergalactic
medium (IGM) with quasar absorption lines.  The \Lya forest in
particular provides a valuable probe of the ionisation state of the
IGM (\citealt{Fan02,Fan06b,Songaila04,Becker07}).  However, the
increasing opacity of the IGM to \Lya photons, culminating in the
appearance of the \cite{GunnPeterson65} trough at $z \simeq 6$
(\citealt{Becker01,White03}), ultimately limits the utility of the
\Lya forest at the highest observable redshifts.  This limitation has
led many authors to analyse the small regions which exhibit
transmission through the \Lya forest even at $z>6$.  These highly
ionised proximity zones lie between the quasar \Lya emission line and
the red-most edge of the Gunn-Peterson trough, and are due to the
enhanced ionisation of hydrogen close to the quasar.

Previous analyses have focused on using these regions to examine the
ionisation state of the IGM with low to moderate resolution spectra
($R\sim 3000-6000$).  These studies have typically modelled the extent
and/or shape of the observed transmission to obtain constraints on the
IGM ionisation state at $z\simeq 6$ ({\it e.g.}
\citealt{Fan06b,MesingerHaiman07,AlvarezAbel07,BoltonHaehnelt07c,Wyithe08,Maselli09}).
However, the proximity zones can also be used to probe the {\it
  thermal state} of the IGM at high redshift.  High resolution ($R\sim
40\,000$) spectra resolve the thermal broadening kernel, enabling the
widths of \Lya absorption lines, and hence the temperature of the gas
in the proximity zones, to be directly measured (\citealt{Becker05}).
The line widths will be sensitive to any photo-heating caused by the
quasar itself (\citealt{MiraldaRees94,BoltonHaehnelt07}).
Furthermore, since the cooling timescale in the low density IGM is
long, information on the reionisation history {\it prior} to any
quasar activity will also be encoded in these absorption lines
(\citealt{Haehnelt98,Theuns02,HuiHaiman03}).

Earlier measurements of the IGM thermal state using the \Lya forest at
$2\leq z \leq 4.5$ have indeed yielded valuable insights into the
epoch of \HeII reionisation at $z\simeq 3$
(\citealt{Schaye00,Ricotti00,McDonald01,Zaldarriaga02,Lidz09}).
However, given enough time following reionisation, the IGM temperature
will eventually reach an asymptotic value which depends only on the
spectral shape of the ionising background.  Information on the earlier
IGM thermal evolution, and hence the \HI reionisation history at high
redshift, will then be effectively erased
(\citealt{Theuns02,HuiHaiman03}).  Consequently, it is desirable to
study the thermal state of the IGM as close as possible to the \HI
reionisation epoch, where the temperature retains a more recent memory
of \HI photo-heating (\citealt{Cen09,FurlanettoOh09}).  The line
widths in proximity zones at $z\simeq 6$ thus contain valuable
information on the thermal history of the IGM at $z\ga 6$ as well as
the impact of any photo-heating by the quasar itself on its
environment (\citealt{BoltonHaehnelt07,Lidz07}).

In this paper we perform an analysis of \Lya absorption line widths in
the proximity zone of the $z=6$ quasar SDSS~J0818$+$1722.  We compare
high resolution Keck/HIRES data with detailed synthetic spectra
constructed using hydrodynamical simulations and line-of-sight
radiative transfer to obtain the first direct constraint on the IGM
temperature around a quasar at $z\simeq 6$.  As such, this work
represents a first step toward developing a procedure which may be
applied to larger data sets at high redshift in the future.  We begin
in section 2 by introducing the observational data and numerical
simulations we use for our analysis.  We discuss our methodology in
section 3, and examine the systematic uncertainties which may impact
on our results in section 4.  We present our temperature measurement
in section 5.  In section 6 we consider the implications of our
results for the IGM reionisation history around SDSS~J0818$+$1722
before finally concluding in section 7.  All distances are expressed
in comoving units unless otherwise stated.

%%%%%%%%%%%%%%%%%%%%%%%%%%%%%%%%%%%%%%%%%%%%%%%%%%%%%%%%%%%%%%%%%%%%%	
%%%%%%%%%%%%%%%%%%%%%%%%%% SECTION 2 %%%%%%%%%%%%%%%%%%%%%%%%%%%%%%%%
%%%%%%%%%%%%%%%%%%%%%%%%%%%%%%%%%%%%%%%%%%%%%%%%%%%%%%%%%%%%%%%%%%%%%

\section{Data and numerical modelling}
\subsection{Observational data}

Our temperature measurements are based on a high-resolution spectrum
of the $z=6.00$ quasar SDSS~J0818$+$1722 (\citealt{Fan06b}).  Keck
HIRES data were taken in February 2006 using the upgraded detector.
We employed a 0.86\arcsec slit, which produces a resolution of $R =
40\,000$ (FWHM = $6.7\rm\,km\,s^{-1}$).  The total integration time
was 7.5 hours.  The raw data were processed using a custom set of IDL
routines that includes optimal sky subtraction (\citealt{Kelson03}).
In order to achieve the highest possible signal-to-noise ($S/N$)
ratio, the final one-dimensional spectrum was optimally extracted from
all exposures simultaneously after relative flux calibrations were
applied to remove the blaze functions from individual orders.  This
allowed us to efficiently reject cosmic rays and other bad pixels
while preserving as many counts from the object as possible.  The
final $S/N$ near the \Lya emission line was $\sim 15$ per
2.1~km~s$^{-1}$ binned pixel.

Continuum fitting in the proximity zone region of a $z \sim 6$ quasar
is challenging due to the strong absorption present in the blue side
of the \Lya emission line.  We first divided the combined spectrum by
a power-law with $F_{\nu} \propto \nu^{-0.5}$, normalised near
$1280\,(1+z)$~\AA.  The value of the power-law slope only weakly
affects the overall continuum level in the proximity zone.  Next we
fit a slowly varying spline to the emission line.  The blue side of
the line profile was made to roughly mirror the red side, which could
be drawn though the unabsorbed continuum.  This produced a generally
Gaussian shape, which we joined smoothly onto the underlying
power-law.  The continuum was set $\sim 10$ per cent above the tops of
the transmission peaks within $\sim 2000\rm \,km\,s^{-1}$ of the
quasar redshift, after which the peaks were allowed to fall further
below the continuum.  Although the continuum level is obscured by the
high levels of absorption, we estimate that our fit is within $\sim
20$ per cent of the correct value over the wavelength range of
interest.

\subsection{Hydrodynamical simulations}

\begin{table*}
  \centering
  \caption{Hydrodynamical/radiative transfer simulations used in this
    work.  From left to right, the columns list the simulation
    identifier, the box size, the total particle number, the gas
    particle mass, the scaling factors for the UVB photo-heating rates
    (see main text for details), the median volume weighted gas
    temperature at mean density, $T_{0}$, and the power law slope of
    the temperature-density relation, $T=T_{0}\Delta ^{\gamma-1}$,
    both prior to and after \HeII photo-heating by the quasar.  The
    slope is given in terms of $\gamma-1 = 2(\log T_{0} - \log T_{-0.5}$),
    where $T_{-0.5}$ is the median volume weighted gas temperature at
    $\log \Delta =-0.5$.  The temperatures are quoted to three
    significant figures only.  Note a power-law temperature-density
    relation following \HeII photo-heating by the quasar is an
    approximation only (see Figure~\ref{fig:trho}).}
    \begin{tabular}{c|c|c|c|c|c|c|c|c|c}
      \hline
   Model   & $L~[h^{-1}\rm\, Mpc]$ & Particles & $M_{\rm gas}~[h^{-1}M_{\odot}]$ & $\zeta$ & $\xi$ & $T_{0,\rm i}$ [K] & $\gamma_{\rm i}-1$ & $T_{0}$ [K] & $\gamma-1$ \\
   \hline
    A &  10 & $2\times 512^{3}$   &  $9.2\times 10^{4}$ & 0.30 & 0.0 &
    $4\,300$ & 0.39 & $13\,500$ & 0.12 \\
    B &  10 & $2\times 512^{3}$   &  $9.2\times 10^{4}$ & 0.80 & 0.0 &
    $8\,500$ & 0.40 & $17\,500$ & 0.18 \\
    C &  10 & $2\times 512^{3}$   &  $9.2\times 10^{4}$ & 1.45 & 0.0 &
    $13\,000$ & 0.41 & $21\,500$ & 0.22 \\
    D &  10 & $2\times 512^{3}$   &  $9.2\times 10^{4}$ & 2.20 & 0.0 &
    $17\,300$ & 0.41 & $25\,700$ & 0.25 \\
    E &  10 & $2\times 512^{3}$   &  $9.2\times 10^{4}$ & 3.10 & 0.0 &
    $21\,900$ & 0.41  & $29\,900$ & 0.27 \\
    F &  10 & $2\times 512^{3}$   &  $9.2\times 10^{4}$ & 4.20 & 0.0 &
    $26\,800$ & 0.41  & $34\,400$ & 0.28 \\
    G &  10 & $2\times 512^{3}$   &  $9.2\times 10^{4}$ & 5.30 & 0.0 &
    $31\,100$ & 0.40  & $38\,500$ & 0.29  \\
    H &  10 & $2\times 512^{3}$   &  $9.2\times 10^{4}$ & 1.45 & -1.0 &
    $12\,800$ & 0.06 & $21\,400$ & 0.04 \\
    L1 &  10 & $2\times 128^{3}$   &  $5.9\times 10^{6}$ & 1.45 & 0.0 &
    $13\,600$ & 0.40 & $22\,300$ & 0.22 \\
    L2 &  10 & $2\times 256^{3}$   &  $7.3\times 10^{5}$ & 1.45 & 0.0 &
    $13\,200$ & 0.40   & $22\,000$ & 0.21 \\
    L3 &  40 & $2\times 512^{3}$   &  $5.9\times 10^{6}$ & 1.45 & 0.0 &
    $13\,400$ & 0.41 & $20\,500$ & 0.23 \\
    W &  10 & $2\times 512^{3}$   &  $9.2\times 10^{4}$ & 1.45 & 0.0 &
    $13\,000$ & 0.41  & $21\,700$ & 0.22  \\
    \hline
\end{tabular}
\label{tab:sims}

\end{table*}

The synthetic proximity zone spectra used in this study are
constructed using line-of-sight density, velocity and temperature
fields drawn from cosmological hydrodynamical simulations.  The
simulations were performed using a customised version of the parallel
Tree-SPH code {\small GADGET-3}, which is an updated version of the
publicly available code {\small GADGET-2} (\citealt{Springel05}).  We
use twelve different hydrodynamical simulations in this work,
summarised in Table~\ref{tab:sims}.  Nine simulations (A-H and W) were
performed in $10h^{-1}$ Mpc periodic boxes containing $2 \times
512^{3}$ gas and dark matter particles.  The simulations are
specifically designed to resolve the \Lya forest at high redshift
(\citealt{BoltonBecker09}).  However, they employ a relatively small
box size to achieve the required mass resolution.  In order to assess
the effect of box size on our results, we performed two further
simulations with identical mass resolution but different box sizes of
$10h^{-1}$ Mpc (L1) and $40h^{-1}$ Mpc (L3), respectively.  Finally,
another $10h^{-1}$ Mpc simulation, L2, was performed with $2 \times
256^{3}$ gas and dark matter particles to provide an additional test
of convergence with mass resolution.

The simulations were all started at $z=99$, with initial conditions
generated using the transfer function of \cite{EisensteinHu99}.  The
cosmological parameters are $\Omega_{\rm m}=0.26$,
$\Omega_{\Lambda}=0.74$, $\Omega_{\rm b}h^{2}=0.023$, $h=0.72$,
$\sigma_{8}=0.80$, $n_{\rm s}=0.96$, consistent with recent studies of
the cosmic microwave background (\citealt{Komatsu09,Reichardt09}).
The IGM is assumed to be of primordial composition with a helium
fraction by mass of $Y=0.24$ (\citealt{OliveSkillman04}).  The
gravitational softening length was set to $1/30^{\rm th}$ of the mean
linear interparticle spacing and -- with the exception of model W --
star formation was included using a simplified prescription which
converts all gas particles with overdensity $\Delta = \rho/\langle
\rho \rangle > 10^{3}$ and temperature $T<10^{5}\rm~K$ into
collisionless stars.  Model W instead uses the multi-phase star
formation and feedback model of \cite{SpringelHernquist03}.  We use
this to explore the impact of strong galactic winds on our results.

The gas in the simulations is assumed to be optically thin and in
ionisation equilibrium with a spatially uniform ultraviolet background
(UVB).  The UVB corresponds to the galaxies and quasars emission model
of \cite{HaardtMadau01}.  Hydrogen is reionised at $z=9$ and gas with
$\Delta \la 10$ subsequently follows a tight power-law
temperature-density relation, $T=T_{0}\Delta^{\gamma-1}$, where
$T_{0}$ is the temperature of the IGM at mean density
(\citealt{HuiGnedin97,Valageas02}).  In order to explore a variety of
thermal histories, we rescale the \cite{HaardtMadau01} photo-heating
rates by different constants in models A-H.  In each simulation we
assume $\epsilon_{i}=\zeta \Delta^{\xi}\epsilon_{i}^{\rm HM01}$, where
$\epsilon_{\rm i}^{\rm HM01}$ are the \cite{HaardtMadau01}
photo-heating rates for species $i=[{\rm H \,\scriptstyle I},{\rm He
    \,\scriptstyle I}, {\rm He \,\scriptstyle II}]$ and $\zeta$, $\xi$
are constants listed in Table 1.  The L1, L2, L3 and W simulations use
same thermal history as model C.  A density dependent rescaling of the
photo-heating rates has been applied to model H, giving $\gamma \simeq
1$ while maintaining a similar $T_{0}$ to model C.  All the other
hydrodynamical simulations in this study have $\gamma \simeq 1.4$.
Simulation outputs were obtained from each model at $z=6.01$.  The
values of $T_{0}$ and $\gamma$ in the hydrodynamical simulations are
listed in columns 7 and 8 of Table 1.

\subsection{Line-of-sight radiative transfer models}

In order to correctly model photo-ionisation and heating\footnote{The
  ionisation and thermal state of the IGM is fully recomputed in
  post-processing by our radiative transfer algorithm; we only use the
  hydrodynamical simulations for the initial conditions (gas density,
  peculiar velocity and temperature) of the gas.  The gas overdensity
  and peculiar velocity field thus remain static in the RT
  calculation, and the hydrodynamical response of the gas to the
  photo-heating induced by the quasar is not modelled.  This effect
  should be small over the quasar lifetime we consider ($t_{\rm
    Q}\simeq 10^{7}\rm\,yr$, {\it cf.} the sound crossing timescale
  $t_{\rm sc} \simeq 6.7 \times 10^{8}(L/10 \rm\,
  proper\,kpc)(T/10^{4}\rm\,K)^{-1/2}\rm\,yr$ and the dynamical
  timescale $t_{\rm dyn}\simeq 4.2\times
  10^{9}\Delta^{-1/2}[(1+z)/7]^{-3/2}\rm\,yr$).}  by the quasar, we
have implemented a line-of-sight radiative transfer (RT) scheme in the
hydrodynamical simulations to account for the propagation of ionising
radiation into the IGM.  For this we use the multi-frequency
line-of-sight radiative transfer algorithm described by
\cite{BoltonHaehnelt07} and updated in \cite{Bolton09}.

We follow the procedure outlined in \cite{BoltonHaehnelt07} for
constructing the initial conditions for our  simulations
including radiative transfer. We use a friends-of-friends halo finding 
algorithm with a linking length of $0.2$ to identify haloes in the
hydrodynamical simulations.
We then select the ten most massive haloes in each simulation and
extract lines-of-sight in different orientations around them.  The
exact halo mass has little impact; we find it is
more important to resolve the IGM along the line-of-sight rather than
the host halo itself.  We discuss this issue further in
section~\ref{sec:sys}.  The halo lines-of-sight are then spliced 
with lines-of-sight drawn at random from the rest of the simulation,
resulting in 30 different lines-of-sight $55h^{-1}$ Mpc in length, all
of which start at the location of a halo.

For each line-of-sight we compute the transfer of ionising radiation
through the IGM from a quasar which emits $\dot N$ photons per second
above the \HI ionisation threshold,

\begin{equation} \dot{N} = \int_{\nu_{\rm HI}}^{\infty}
  \frac{L_{\nu}}{h_{\rm p}\nu}\, d\nu, \label{eq:Ndot} \end{equation}

\noindent
with a generic broken power law spectrum given by

\begin{equation}
L_{\nu}\propto \cases{\nu^{-0.5} &($1050<\lambda < 1450\,$\AA),\cr
  \noalign{\vskip3pt}\nu^{-1.5} &($\lambda<1050\,$\AA).\cr}
\end{equation}

\noindent
This spectrum has an extreme UV (EUV) spectral index $\alpha_{\rm
  s}=1.5$, consistent with radio quiet quasars at lower redshift
(\citealt{Telfer02}).  Note, however, the exact spectrum and hence
ionising luminosity of the quasar is rather uncertain at $z=6$.
Adopting a harder (softer) EUV spectral index will increase (decrease)
the ionising luminosity and the amount of photo-heating around the
quasar.  We assume $M_{1450}=-27.4$ for the absolute magnitude of the
quasar, corresponding to ${\dot N}=2.74 \times 10^{57}\rm\, s^{-1}$
for our chosen spectrum.  This magnitude corresponds to the $M_{1450}$
obtained for SDSS~J0818$+$1722 (\citealt{Fan06b}).  Lastly, we use a
duration of $t_{\rm Q}=10^{7}\rm\, yr$ for the phase when the quasar is
optically bright (\citealt{Haehnelt98b,Martini04,Croton09}).  This
parameter is also somewhat uncertain; we discuss the implications of
this in more detail shortly.

In all our RT simulations, we initialise the ionisation state of the
IGM by assuming photo-ionisation equilibrium with a UVB given by

\begin{equation} J_{\nu} = J_{-21} \left( \frac{\nu}{\nu_{\rm HI}}
\right)^{-3} \times \cases{1 & ($\nu_{\rm HI} \leq \nu <
  \nu_{\rm HeII}$),\cr \noalign{\vskip3pt} 0 & ($\nu_{\rm
    HeII} \leq \nu$),\cr}
\label{eq:UVB}
\end{equation}

\noindent
where $J_{-21}$ is the amplitude of the UVB at the \HI ionisation
threshold in proper units of
$10^{-21}\rm\,erg\,s^{-1}\,cm^{-2}\,Hz^{-1}\,sr^{-1}$.  We adopt a
soft, stellar like spectrum and assume \HI and \HeI is already highly
ionised by $z\leq 6$ ({\it e.g.}
\citealt{ChoudhuryFerrara06,Becker07,Pritchard09}).  There is no
evidence to suggest the hydrogen in the IGM surrounding
SDSS~J0818$+$1722 was significantly neutral prior to the quasar
turning on (see \citealt{BoltonHaehnelt07} for a detailed discussion
of this point).  The exact choice of UVB spectral index at frequencies
below the \HeII ionisation threshold is unimportant, since the IGM
temperature is initialised using the hydrodynamical simulations.

\begin{figure}
\begin{center}
  \includegraphics[width=0.45\textwidth]{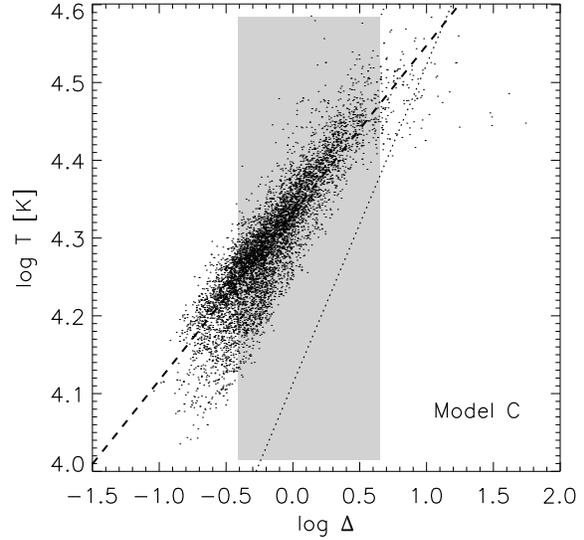}
\vspace{-0.3cm}
\caption{Scatter plot of the volume weighted temperature-density plane
  for the 30 lines-of-sight used in model C.  A random subset of
  pixels $250\rm\,km\,s^{-1}\leq v_{\rm H} \leq 3250\rm\,km\,s^{-1}$
  from the quasar redshift are displayed for clarity of presentation.
  The dashed line shows the final (approximate) power-law
  temperature-density relation after \HeII photo-heating by the
  quasar, while the dotted line corresponds to the initial IGM
  temperature-density relation (taken from the hydrodynamical
  simulation) before the quasar turned on.  There is significant
  scatter in the relationship between temperature and density
  following quasar activity.  The shaded region corresponds to $95$
  per cent of all pixels with normalised flux $0.05\leq F \leq 0.95$
  in the corresponding proximity zone spectra.  This gives an
  indication of the densities which transmission in the proximity zone
  is sensitive to (see the discussion in section~\ref{sec:sys2}).}
\label{fig:trho}
\end{center}
\end{figure}

However, \HeII reionisation, which must be driven by sources with hard
ionising spectra, is thought to be delayed until around $z\simeq 3$
when the number density of quasars begins to peak
(\citealt{MadauMeiksin94,FurlanettoOh08b}).  We model this delay by
truncating the UVB above the \HeII ionisation threshold.  When the
quasar turns on in our RT models, its hard ionising spectrum will
therefore reionise and photo-heat the \HeII in the surrounding IGM,
resulting in the so-called ``thermal proximity effect''
(\citealt{MiraldaRees94,Theuns02c,Meiksin09b}).  The main effect of
\HeII reionisation by the quasar is to flatten the temperature density
relation and boost the temperature at mean density by $\sim 7000-
9000\rm\,K$ within the proximity zone.  The volume weighted
temperature-density plane from one of our RT simulation sets (model C)
is displayed in Figure~\ref{fig:trho}.  The final (approximate)
temperature-density relation in the RT simulations after \HeII
photo-heating by the quasar are given in the last two columns of
Table~\ref{tab:sims}.  As already discussed, the exact amount of
heating depends on the assumed EUV spectral index.  Note also that if
the quasar lifetime is significantly shorter (longer) than the generic
value of $t_{\rm Q}=10^{7}\rm\, yr$ we have assumed, the extent of the
region where \HeII is photo-ionised and heated by the quasar will be
smaller (larger).  In this work we consider the IGM within
$33\rm\,Mpc$ of SDSS~J0818$+$1722, a scale over which \HeII is fully
ionised for $t_{\rm Q}=10^{7}\, \rm yr$ for our choice of quasar
spectrum.  Longer quasar lifetimes will make little difference to our
results.  However, if $t_{\rm Q} \la 10^{7}\rm\, yr$ our models will
overestimate the size of the \HeIII region (and hence \HeII
photo-heating) around the quasar.

\begin{figure*}
\centering
\begin{minipage}{180mm}
\begin{center}
\psfig{figure=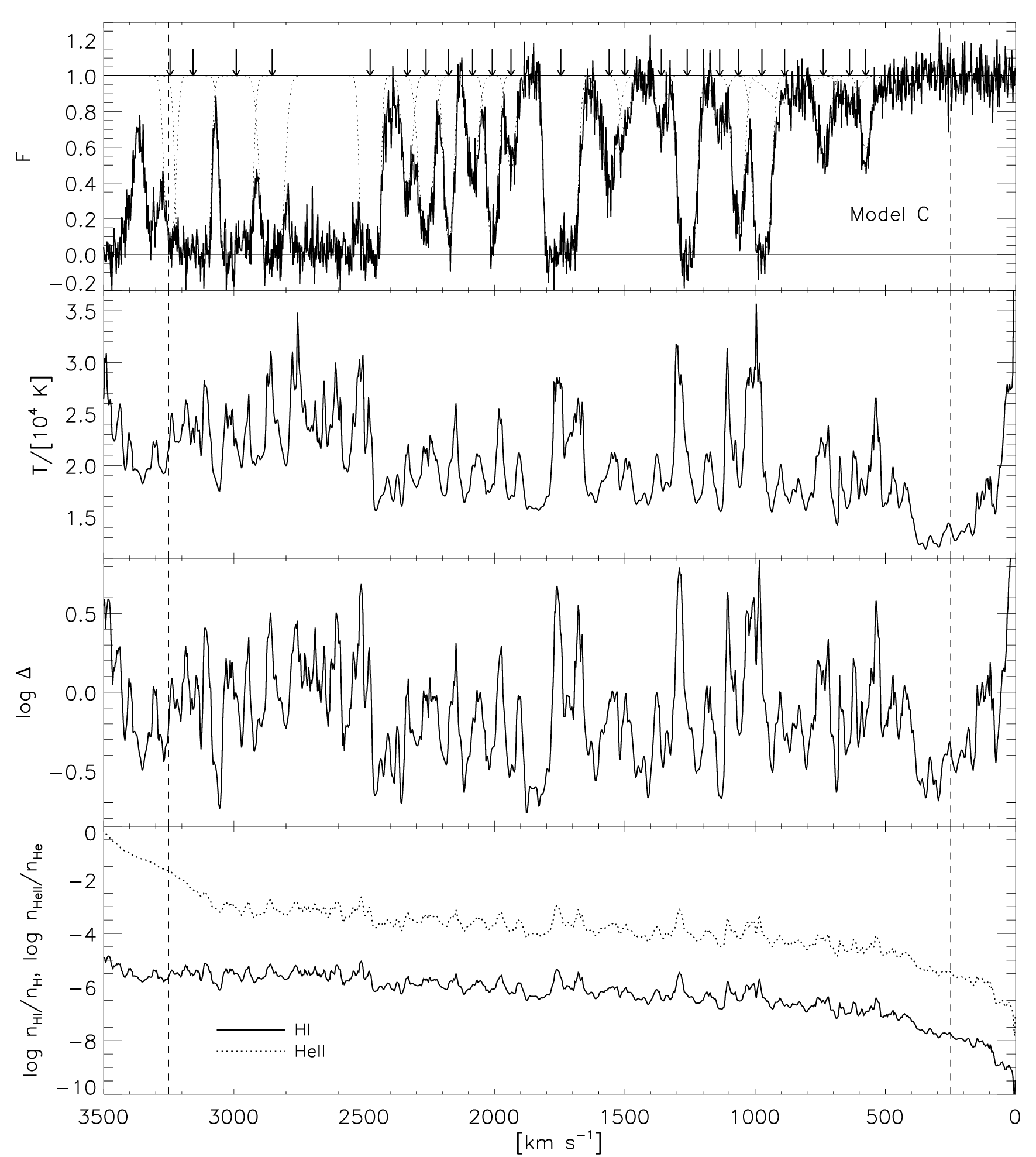,width=0.75\textwidth}
\psfig{figure=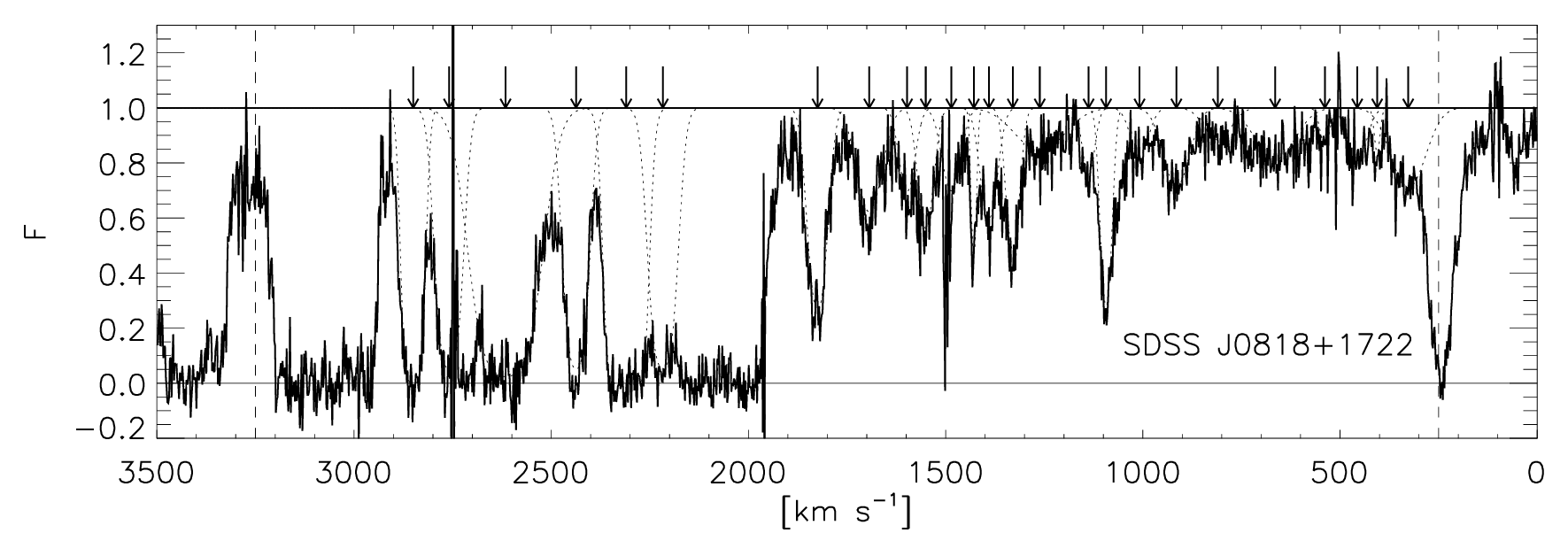 ,width=0.75\textwidth}
\caption{{\it Upper four panels:} The topmost panel shows an example
  synthetic spectrum at $z=6$ drawn from model C.  The quasar is
  situated at $v_{\rm H}=0\rm \,km\,s^{-1}$ on the right hand side of
  each panel.  The vertical dashed lines indicate the velocity range
  over which Voigt profiles are fitted to the data,
  $250\rm\,km\,s^{-1}\leq v_{\rm H} \leq 3250\rm\,km\,s^{-1}$ ($2.6\rm
  \,Mpc \leq R \leq 33.3\, Mpc$).  The vertical arrows mark the
  redshift of the Voigt profile fits made with VPFIT, while the fits
  themselves are shown by the dotted curves. The corresponding gas
  temperature, overdensity and fractional abundances of \HI and \HeII
  along the line-of-sight are shown in the subsequent three panels.
  {\it Lower panel:} Keck/HIRES spectrum of SDSS J0818+1722 at
  $z=6.00$. }
\label{fig:spectrum}
\end{center}
\end{minipage}
\end{figure*}

\begin{figure*}
\centering
\begin{minipage}{180mm}
\begin{center}
\psfig{figure=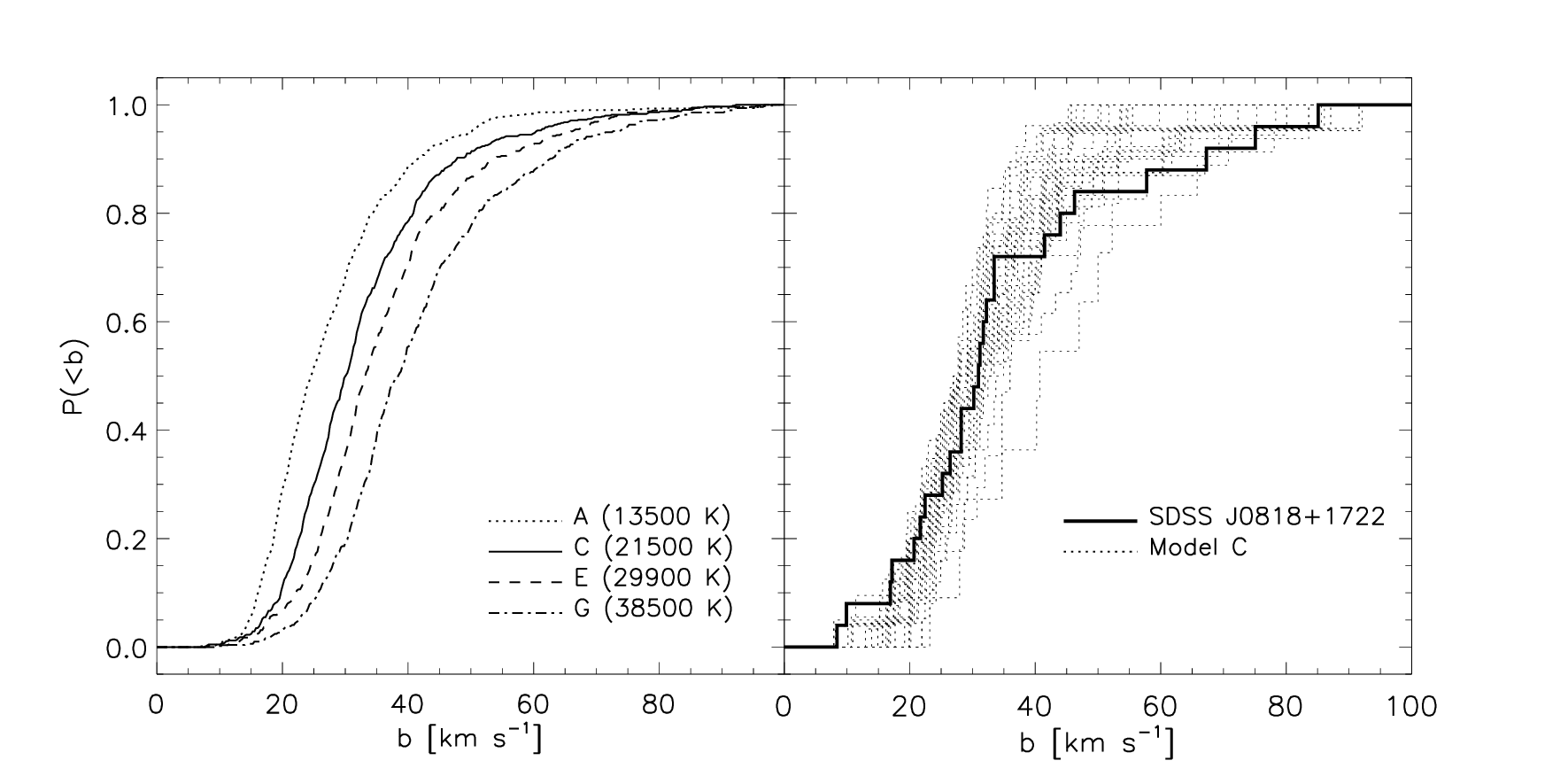,width=0.85\textwidth}

\vspace{-0.3cm}
\caption{{\it Left:} The Doppler parameter CPDF for four of our RT
  models.  The simulated CPDFs correspond to the data from all 30
  lines-of-sight in each model.  The CPDF is clearly sensitive to the
  gas temperature in the quasar proximity zone, with the CPDF shifting
  to higher b-values for increasing temperature.  The median Doppler
  parameters for the simulated models displayed are $b_{\rm
    med}=(24.5,30.0,33.2,38.4)\rm\,km\,s^{-1}$.  {\it Right:} The
  observed distribution for SDSS~J0818$+$1722 (thick solid curve),
  which has $b_{\rm med}=31.0\rm\,km\,s^{-1}$, in comparison to the
  CPDFs for the individual lines-of-sight in model C.  This gives an
  indication of the scatter in the CPDF from one line-of-sight to the
  next.}
\label{fig:btemp}
\end{center}
\end{minipage}
\end{figure*}

Lastly, the amplitude of the UVB at the \HI ionisation edge in
Eq.~(\ref{eq:UVB}), $J_{-21}$, is set using the density dependent
model presented in \cite{Wyithe08}.  Briefly, this model accounts for
the bias in the overdense region around the quasar host halo and is
computed as a function of the proper time along the trajectory of a
photon emitted by the quasar.  This results in an enhanced
contribution to the UVB near the quasar host halo relative to the mean
IGM.  The latter is calibrated to match measurements of the \HI
photo-ionisation rate derived from the observed \Lya forest opacity at
$4\leq z \leq 6$ (\citealt{BoltonHaehnelt07b}).  A detailed
description of the model and its application to high redshift \Lya
forest spectra may be found in \cite{Wyithe08}.  Note that the UVB
amplitude is important for the observed {\it size} of a proximity zone
(\citealt{Wyithe08}), but the precise value of this parameter is not
important for the gas temperatures around the quasar we explore in
this work.

\subsection{Spectra construction} \label{sec:makespec}

We follow standard procedure to construct synthetic proximity zone
spectra from the output of each of our RT models ({\it e.g}
\citealt{Theuns98}).  The spectra are then convolved with a Gaussian
instrument profile with $\rm FWHM=6.7\,km\,s^{-1}$, resampled onto
pixels of width $2.1\rm\,km\,s^{-1}$ and Gaussian distributed noise
with $S/N=12$ is added.  These values are chosen to match the observed
spectrum of SDSS~J0818$+$1722.  To account for uncertainties in the
continuum placement on the observed data, we renormalise the synthetic
data by the highest flux in $1000\rm\,km\,s^{-1}$ segments along each
line-of-sight.  Although this will not account for continuum
uncertainties perfectly, it will approximate the way the continuum is
fitted to the real data and should reduce a potential systematic bias.
We consider this issue further in section~\ref{sec:sys2}.  Lastly, we
rescale the optical depths in each pixel of the synthetic spectra by a
constant to match the mean flux of the observed data, $\langle F
\rangle_{\rm obs}=I_{\rm obs}/I_{\rm cont}=0.486$, within the range
$250\rm\,km\,s^{-1} \leq v_{\rm H}\leq 3250\rm\,km\,s^{-1}$ ($2.6\rm
\,Mpc \leq R \leq 33.3\, Mpc$).  An example spectrum drawn from model
C is displayed in Figure~\ref{fig:spectrum} along with the observed
line-of-sight we use in this work, SDSS~J0818$+$1722.

%%%%%%%%%%%%%%%%%%%%%%%%%%%%%%%%%%%%%%%%%%%%%%%%%%%%%%%%%%%%%%%%%%%%%	
%%%%%%%%%%%%%%%%%%%%%%%%%% SECTION 3 %%%%%%%%%%%%%%%%%%%%%%%%%%%%%%%%
%%%%%%%%%%%%%%%%%%%%%%%%%%%%%%%%%%%%%%%%%%%%%%%%%%%%%%%%%%%%%%%%%%%%%

\section{Analysis procedure}\label{sec:analysis}
\subsection{Voigt profile fitting}
 
In this work we use the cumulative probability distribution function
(CPDF) of the Doppler widths of \Lya absorption lines, $b$, in the
proximity zone as our probe of the IGM temperature.  The CPDF has the
advantage of fully using the limited data available (our analysis of
SDSS~J0818$+$1722 yields only $25$ Doppler widths) and it avoids
binning and the associated loss of information.  Although the
absorption lines will not all be purely thermally broadened
(\citealt{Theuns00}), all absorption lines will nevertheless be
smoothed on a scale associated with the thermal broadening kernel.
The entire CPDF is therefore sensitive to the IGM temperature.
However, the lack of a direct relationship between the Doppler
parameters and temperature means detailed synthetic spectra are
crucial for calibrating this statistic.

We obtain the CPDF for both the observed and synthetic data via Voigt
profile fitting.  We perform the analysis using an automated version
of the Voigt profile fitting package
VPFIT.\footnote{http://www.ast.cam.ac.uk/$\sim$rfc/vpfit.html} We
choose to include lines in the range $250\rm\,km\,s^{-1} \leq v_{\rm
  H}\leq 3250\rm\,km\,s^{-1}$ ($2.6\rm \,Mpc \leq R \leq 33.3\, Mpc$)
from the quasar redshift only.  The lower limit is chosen to avoid
edge effects in the synthetic data, while beyond $3250\rm\,km\,s^{-1}$
it becomes impossible to fit Voigt profiles reliably due to the
increasing opacity of the IGM.  All lines with relative errors in
excess of $50$ per cent are rejected from our final analysis
(\citealt{Schaye99}).  This removes most of the very narrow lines
which tend to be in blends. We also discard all lines with Doppler
parameters $b>100\rm\,km\,s^{-1}$ and column densities $\log N_{\rm
  HI}>17$.  Broad shallow lines with $b>100\rm\,km\,s^{-1}$ are almost
always added near the continuum level to improve the overall fit,
while strong lines with damping wings tend to be added near the edge
of the proximity zone where there is very little transmission.  These
cuts remove 32 and 25 per cent of identified Voigt profiles in the
observed and simulated data sets, respectively.  Lastly, we note that
some ambiguous fits will still remain in our analysis.  However, since
we treat the observed and synthetic spectra in exactly the same way we
expect any potential bias to be small.

The sensitivity of the resulting Doppler width CPDF to the gas
temperature is demonstrated in the left panel of
Figure~\ref{fig:btemp}, where the CPDFs for four of our RT simulation
sets are displayed.  The CPDFs correspond to the data from all 30
lines-of-sight in each RT simulation set.  It is clear the entire CPDF
is sensitive to the gas temperature in the proximity zone.  Increasing
the temperature of the IGM shifts the CPDF towards higher b-parameter
values, increasing the median Doppler parameter.  The median Doppler
parameters for the models displayed, which span a range of
$13\,500\rm\,K\leq T_{0} \leq 38\,500\,K$, are $b_{\rm
  med}=(24.5,30.0,33.2,38.4)\rm\,km\,s^{-1}$.  The observed
distribution, which has $b_{\rm med}=31.0\rm\,km\,s^{-1}$, is
displayed as the thick solid curve in the right panel of
Figure~\ref{fig:btemp} along with the CPDFs for the {\it individual}
lines-of-sight in model C.  This gives an indication of the scatter in
the CPDF from one line-of-sight to the next for the same underlying
model.  A two distribution Kolmogorov-Smirnov (KS) test
(\citealt{Press92}) using the average (all 30 lines-of-sight) and
observed CPDFs yields $P_{\rm KS}=0.06,\,0.94,\,0.21$ and $4\times
10^{-4}$ for models A, C, E and G respectively.  This does not rule
out the null hypothesis that the observed CPDF is drawn from the same
distribution as models A, C and E.

\subsection{Monte-Carlo analysis}

\begin{figure*}
\centering
\begin{minipage}{180mm}
\begin{center}
\psfig{figure=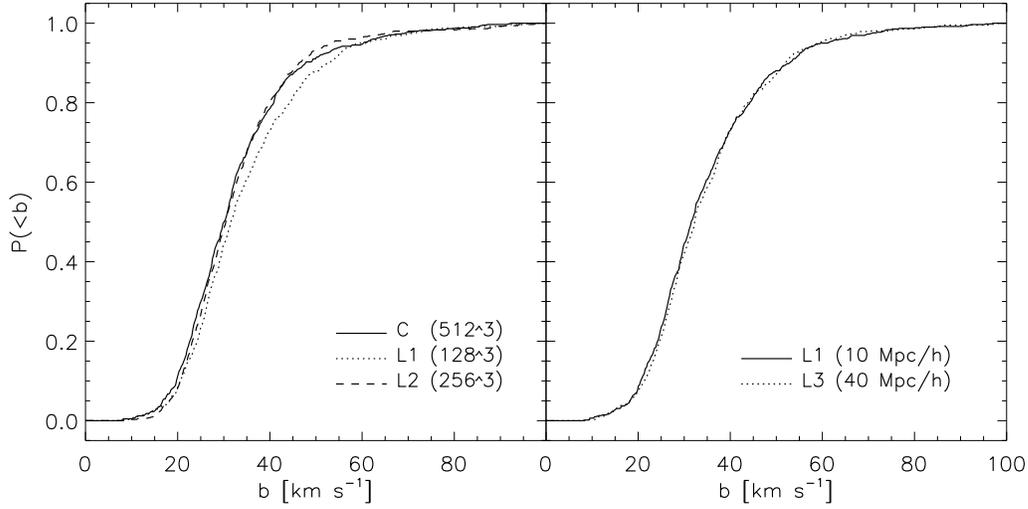,width=0.85\textwidth}

\vspace{-0.3cm}
\caption{Numerical convergence tests for the simulated Doppler width
  CPDF.  The CPDFs correspond to the data from all 30 lines-of-sight
  in each RT simulation set.  {\it Left:} Test of the hydrodynamical
  simulation mass resolution.  Although the CPDF from the lowest
  resolution L1 model has not converged, the distribution is well
  captured for our fiducial mass resolution of $9.2\times
  10^{4}h^{-1}M_{\odot}$ (model C) {\it Right:} Test of the simulation
  box size.  The CPDF is well converged, indicating our choice of a
  $10h^{-1}$ Mpc box is sufficient.}
\label{fig:bsysn}
\end{center}
\end{minipage}
\end{figure*}

We now turn to describing the methodology which lies at the heart of
our analysis procedure.  We use our synthetic spectra to construct
Monte Carlo realisations of the Doppler parameter CPDF for a range of
models with different IGM temperatures.  To quantify the amount of
scatter in the simulated CPDFs, we use a ``D-statistic'' that is very
similar to the parameter used in a KS test.  For each line-of-sight in
a given simulation set, the D-statistic is the maximum difference
between the Doppler parameter CPDF for that line-of-sight and the CPDF
for all 30 lines-of-sight, such that

\begin{equation} D_{\rm i}= {\rm max} | P(<b)_{\rm i} - P(<b)_{\rm all}|, \,i=1...30,\end{equation}

\noindent
where we preserve the sign of the difference.  The D-statistic CPDF
for a model with known temperature $T_{0}$, $P(<D|T_{0})$, can then be
constructed.  If the D-statistic for an observed line-of-sight,
$D_{\rm obs}$, is known, the D-statistic CPDF may then be used in
combination with Bayes theorem to infer a confidence interval for the
{\it observed} $T_{0}$.  The cumulative probability for the observed
temperature at mean density $T_{0}$ given $D_{\rm obs}$ is:

\begin{equation} P(<T_{0}|D_{\rm obs}) = \int_{0}^{T_{0}}\frac{dP(<T_{0}^{\prime}|D_{\rm obs})}{dT_{0}^{\prime}}dT_{0}^{\prime}.  \end{equation}

\noindent
An application of Bayes theorem, $p(T_{0}|D_{\rm obs}) \propto
p(D_{\rm obs}|T_{0})p(T_{0})$, then leads to

\begin{equation} P(<T_{0}|D_{\rm obs}) = K\int_{0}^{T_{0}}  \frac{dP(<D_{\rm obs}|T_{0}^{\prime})}{dD} \frac{dP(<T_{0}^{\prime})}{dT_{0}^{\prime}} dT_{0}^{\prime}, \label{eq:Tdist}  \end{equation}

\noindent
where $K$ is a constant which normalises the total probability to
unity and $p(T_{0})$ is the prior on $T_{0}$.  In order to obtain our
constraint on $T_{0}$, we evaluate the derivative of the D-statistic
CPDF derived from our simulations at $D_{\rm obs}$ in
Eq.~(\ref{eq:Tdist}).  We adopt a flat prior, $p(T_{0})$, over the
range of temperatures explored in the simulations.  These are set by
the combined effect of the initial temperature of the hydrodynamical
simulations and the subsequent heat input from \HeII photo-heating by
the quasar, such that $13\,500 \rm\,K \leq T_{0} \leq 38\,500\rm\,K$.
This is intended to represent a reasonable range for the IGM
temperature following \HI and \HeII reionisation, and is consistent
with temperatures predicted in simulations of \HeII reionisation
(\citealt{McQuinn09}).

The main drawback of our method is the computational expense of
constructing and analysing synthetic spectra.  We have seven
hydrodynamical simulations with different values of $T_{0}$, so we may
only evaluate $\frac{dP(<D_{\rm obs}|T_{0})}{dD}$ at seven discrete
points.  We use a cubic spline to interpolate between these to obtain
a continuous distribution for integration.  With the current method,
finer sampling of the distribution would require running and analysing
additional simulations.  A less expensive approach would be to use a
single hydrodynamical simulation and simply impose a range of initial
temperature-density relations on the IGM.  However, this would
decouple the initial gas temperature from the gas hydrodynamics and
incorrectly model the effect of pressure (Jeans) smoothing on the gas
distribution ({\it e.g.}  \citealt{Pawlik09,Peeples09}).  Furthermore,
there are only 30 line-of-sight in each simulation set, which makes
the derivative of the D-statistic CPDF somewhat noisy.  This could be
remedied by analysing more lines-of-sight, although the Voigt profile
analysis procedure is time consuming.  For this first attempt we have
instead elected to perform a easily manageable number of simulations.
We therefore smooth the D-statistic CPDF with a Gaussian filter of
width $\sigma=0.025$ before computing its derivative.  We find this to
be the optimum smoothing width, and we have verified that alternative
choices of $\sigma=0.01$ and $0.05$ do not significantly alter our
final results.

On the other hand, the major advantage of our approach is that it
avoids binning the data and is non-parametric.  However, our method
implicitly assumes that the synthetic spectra accurately represent the
observed data.  If our synthetic models are significantly in error,
any temperature constraints will be unreliable.  For this reason, we
now turn to consider numerical convergence and some of the possible
systematic errors which may affect our analysis.

%%%%%%%%%%%%%%%%%%%%%%%%%%%%%%%%%%%%%%%%%%%%%%%%%%%%%%%%%%%%%%%%%%%%%	
%%%%%%%%%%%%%%%%%%%%%%%%%% SECTION 4 %%%%%%%%%%%%%%%%%%%%%%%%%%%%%%%%
%%%%%%%%%%%%%%%%%%%%%%%%%%%%%%%%%%%%%%%%%%%%%%%%%%%%%%%%%%%%%%%%%%%%%

\section{Convergence and systematics}

\subsection{Numerical convergence} \label{sec:sys}

We first consider the effect of mass resolution and box size on the
simulated Doppler width CPDF in Figure~\ref{fig:bsysn}.  The left
panel displays the impact of mass resolution on the CPDF, while the
right panel indicates the effect of box size.  The CPDF is well
converged for our fiducial mass resolution ($9.2 \times
10^{4}h^{-1}\rm \, M_{\odot}$), although note our lowest resolution
model (L1, $5.9 \times 10^{6}h^{-1}\rm \,M_{\odot}$) does not
correctly resolve the line widths.  It is also encouraging that the
box size, and hence the amount of large scale power in our
simulations, has little effect on the CPDF.  The maximum halo mass in
the $40h^{-1}$ Mpc box is $4.33\times 10^{11}h^{-1}\rm\,M_{\odot}$,
which is an order of magnitude greater than the largest halo in our
fiducial simulation volume of $10h^{-1}$ Mpc ($6.13\times
10^{10}h^{-1}\rm\,M_{\odot}$).  However, this has little effect on the
CPDF because we are measuring Doppler parameters in the ambient IGM
rather than in the host environment of the quasar.  Note, however,
that the maximum halo mass we consider is still smaller than the
expected host masses of $\sim 10^{12}-10^{13}\rm\,M_{\odot}$ for
bright $z\simeq 6$ quasars (\citealt{Walter04,VolonteriRees06,Li07}).
We nevertheless find it is much more important to resolve absorption
line widths correctly in our models using a high resolution simulation
rather than using a larger volume.  We conclude that our simulations
are well converged with box size and mass resolution for the purpose
of this study; any differences in the CPDF are small compared to the
contrast between models with different temperatures.

\subsection{Systematic uncertainties} \label{sec:sys2}

\subsubsection{The mean flux}

\begin{figure*}
\centering
\begin{minipage}{180mm}
\begin{center}
\psfig{figure=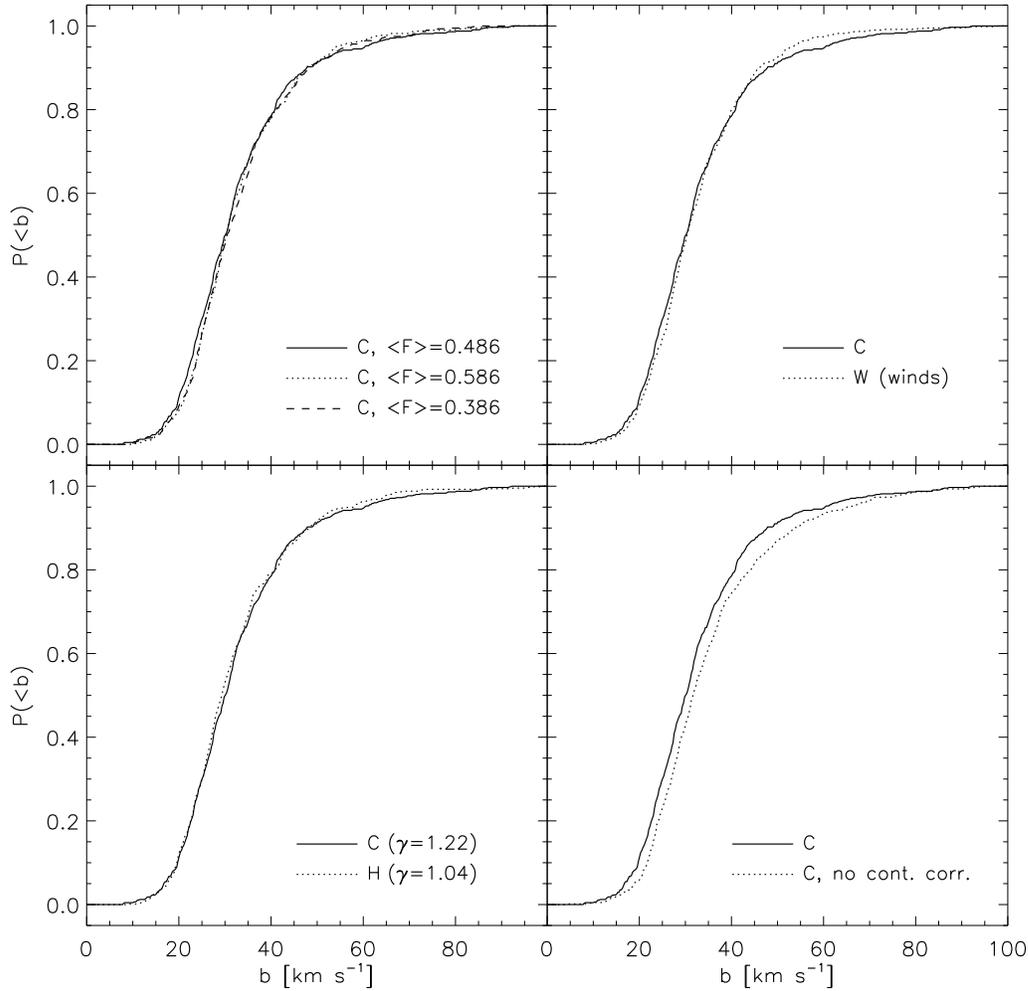,width=0.85\textwidth}
\vspace{-1.5cm}
\caption{Systematic uncertainties and their effect on the Doppler
  parameter CPDF.  {\it Upper left:} The effect of the mean flux
  measured over $250\rm\,km\,s^{-1}\leq v_{\rm H} \leq
  3250\rm\,km\,s^{-1}$.  The fiducial value in this work is $\langle F
  \rangle=0.486$ (solid curve), corresponding to value measured from
  the spectrum of SDSS~J0818$+$1722.  Two further models with $\langle
  F \rangle=0.386$ (dashed curve) and $\langle F \rangle=0.586$
  (dotted curve) are displayed for comparison.  {\it Upper right:} The
  impact of feedback in the form of strong galactic winds, $v_{\rm
    w}=448\rm\,km\,s^{-1}$, implemented using the multi-phase star
  formation model of \citet{SpringelHernquist03}.  Model C (solid
  curve) has no winds while model W (dotted line) is identical to
  model C in all respects aside from the inclusion of galactic winds.
  {\it Lower left:} The effect of the temperature-density relation
  slope, $\gamma$.  Model C (solid curve) has $\gamma \sim 1.2$ while
  model H (dotted curve) has $\gamma\sim 1$ following \HeII
  photo-heating by the quasar.  {\it Lower right:} The effect of
  uncertainty in the continuum placement.  The dotted curve shows the
  CPDF obtained before the standard continuum correction is applied to
  the synthetic data (see main text for details), while the solid
  curve is obtained from the data after correction.  Lowering the
  continuum (it is never raised in the correction) always reduces the
  median Doppler parameter.}
\label{fig:bastro}
\end{center}
\end{minipage}
\end{figure*}

The effect of four different systematic uncertainties on the simulated
Doppler parameter CPDF are displayed in Figure~\ref{fig:bastro}.  The
first we consider is the mean transmitted flux, $\langle F \rangle =
\langle e^{-\tau_{\rm i}} \rangle$, where $\tau_{\rm i}$ is the
optical depth in pixel $i$.  As discussed earlier, when constructing
our synthetic spectra we follow a standard procedure and rescale the
\HI photo-ionisation rate in post-processing by a fixed constant to
produce spectra which match the $\langle F \rangle$ of the observed
data, $\langle F \rangle=0.486$.  It is therefore important to assess
how a different value of $\langle F \rangle$ may influence the Doppler
parameter CPDF.

The effect of the mean flux in the region of the spectra where lines
are fitted, $250\rm\,km\,s^{-1}\leq v_{\rm H} \leq
3250\rm\,km\,s^{-1}$, is shown in the upper left panel of
Figure~\ref{fig:bastro}.  We perform a Voigt profile analysis on two
further models which are identical to model C, except from being
rescaled in post-processing to have $\langle F \rangle=0.386$ and
$\langle F \rangle=0.586$.  It is clear that the mean flux has little
effect on the Doppler parameter CPDF, changing the median Doppler
parameter by less than $0.5\rm\,km\,s^{-1}$ for the examples shown.
This is expected theoretically; as pointed out by
\cite{HuiRutledge99}, the ionising background does not play a role in
setting the {\it shape} of an absorption line -- and hence its Doppler
parameter -- around the peak optical depth.  Note, however, that the
mean flux will indeed be important for the line column densities,
decreasing $N_{\rm HI}$ at fixed density when the photo-ionisation
rate is raised.  We find that uncertainties in the mean flux should
not  be important for this work.

\subsubsection{Galactic winds}

We reach a similar conclusion regarding the impact of galactic winds
on the CPDF in the upper right panel of Figure~\ref{fig:bastro}.  We
have constructed spectra from model W.  This uses the multi-phase star
formation and winds model of \cite{SpringelHernquist03}, but in all
other respects is identical to model C. The wind velocity is $v_{\rm
  w}=448\rm\,km\,s^{-1}$, which is an extreme model in the sense that
it assumes all energy produced by supernovae is converted into kinetic
energy.  The differences in the CPDF are again small, with $b_{\rm
  med}=30.0\rm\,km\,s^{-1}$ and $30.3\rm\,km\,s^{-1}$ for the C and W
models respectively.

\subsubsection{The temperature-density relation}

We do not attempt to measure the density dependence of the IGM
temperature in this work; this is difficult to achieve for a single
quasar spectrum at $z=6$.  Instead we focus on constraining the
temperature at mean density, $T_{0}$.  The assumed power-law slope of
the temperature-density relation, $\gamma$, is therefore a systematic
uncertainty in our analysis.  Note again that a single power-law
temperature-density relation, $T=T_{0}\Delta^{\gamma-1}$, does not
provide an accurate description of the IGM thermal state in our
simulations ({\it e.g.}  Figure~\ref{fig:trho}, and see also
\citealt{Bolton04,Trac08}).  Nevertheless, for the present we find it
convenient to characterise the relationship between temperature and
density as scatter around a power-law.

In the lower left panel of Figure~\ref{fig:bastro} we demonstrate the
effect of $\gamma$ on the Doppler parameter CPDF.  Model C and model H
have $\gamma \sim 1.2$ and $\gamma \sim 1.0$ respectively, with almost
identical temperatures at mean density.  Interestingly, there is very
little difference between the CPDFs, with $b_{\rm
  med}=30.0\rm\,km\,s^{-1}$ and $29.3\rm\,km\,s^{-1}$.  This is the
result one would expect if the CPDF is sensitive to gas densities
around the cosmic mean at $z=6$.  We may test if this is the case by
evaluating the range of optical depth weighted overdensities,
$\Delta_{\tau}=\sum \Delta_{\rm i}\tau_{\rm i}/\sum \tau_{\rm i}$
({\it e.g.} \citealt{Schaye99}), over the range
$250\rm\,km\,s^{-1}\leq v_{\rm H} \leq 3250\rm\,km\,s^{-1}$ in our
simulated spectra.  In Figure~\ref{fig:trho} the shaded region
corresponds to the range of $\Delta_{\tau}$ including 95 per cent of
the pixels with $0.05\leq F \leq 0.95$ in the proximity zone: $0.4
\leq \Delta_{\tau} \leq 4.4$, with a median value of
$\Delta_{\tau}=0.95$.  The transmission in the proximity zone thus
predominantly probes the IGM around mean density, and it is clear why
different values of $\gamma$ have only a small impact on the
CPDF. Future studies at $z\simeq 6$ could utilise higher order Lyman
series absorption sensitive to higher gas densities to obtain
constraints on $\gamma$ (\citealt{Dijkstra04b,FurlanettoOh09}).  For
now, based on these results we shall assume the impact of this
parameter on our analysis is small.

\begin{figure*}
\centering
\begin{minipage}{180mm}
\begin{center}
\psfig{figure=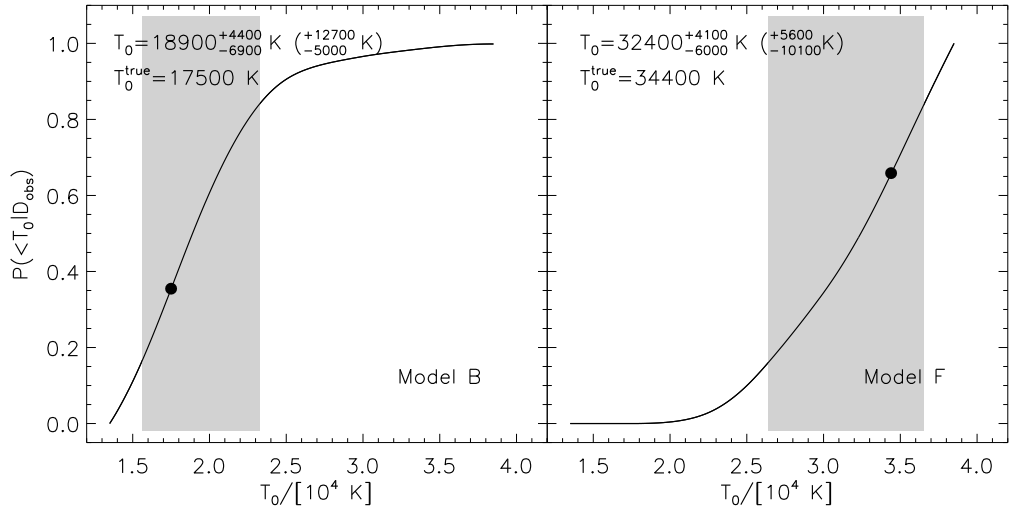,width=0.85\textwidth}
\vspace{-0.3cm}
\caption{Test of our analysis procedure using two random
  lines-of-sight drawn from our simulated data set.  The
  lines-of-sight in the left and right panels are drawn from models B
  and F, respectively.  The solid curves show the cumulative
  distribution for $T_{0}$, $P(<T_{0}|D_{\rm obs})$, computed using
  Eq.~(\ref{eq:Tdist}).  The filled circles show the ``true''
  temperature of the models at mean density, while the shaded regions
  indicate the 68 per cent confidence interval around the recovered
  median.  The numerical values in the top left of each panel
  correspond to the recovered median temperature and the 68 (95) per
  cent confidence intervals.  The large uncertainties are dominated by
  the statistical error associated with measuring the temperature
  along a single line-of-sight.}
\label{fig:Ttest}
\end{center}
\end{minipage}
\end{figure*}

\subsubsection{Continuum placement}

We now  consider the effect of the uncertainty in the intrinsic
emission of the quasar, $I_{\rm cont}$, where $I_{\rm obs}=I_{\rm
  cont}e^{-\tau}$.  At $z=2-3$, the typical uncertainty in the
continuum placement in high resolution \Lya forest data is estimated
to be 1-4 per cent (\citealt{Tytler04,Kirkman05,Kim07,Faucher08}).
However, towards higher redshift, the increasing opacity of the IGM
means the transmitted flux recovers to the unabsorbed level less
and less frequently.  Identifying the continuum level accurately becomes very
challenging, with an estimated uncertainty of up to $\sim 20$ per
cent even for the high-quality spectrum considered here.  
Continuum estimation therefore presents a significant
systematic uncertainty in our analysis.

As it is difficult to identify where the continuum lies on the observed
data at $z\simeq 6$, our approach in this work is instead to treat the
simulated spectra in the same way as the observational data.  The
observational data were normalised using a slowly varying spline fit,
adjusted by hand to roughly follow the profile of the transmission
peaks in the proximity zone.  The continuum correction we apply to our
data, described in section~\ref{sec:makespec}, is designed to mimic
this.  To recap, we renormalise the synthetic data by the highest flux
in $\sim 1000\rm\,km\,s^{-1}$ segments along each line-of-sight.
Although this means we may not correctly identify the ``true''
continuum, in principle this procedure should at least reduce  any
bias (see \citealt{Rauch97,McDonald00} for similar approaches to this
problem at lower redshift).

The lower right panel of Figure~\ref{fig:bastro} displays the effect of
the continuum correction on the CPDF, which always acts to {\it lower}
the continuum on the synthetic spectra.  Although this correction has
little effect near the redshift of the quasar -- the IGM here is very
highly ionised and the flux regularly reaches the unabsorbed level in
our simulations -- it becomes larger towards the blue-most edge of the
region we fit lines to; the \Lya transmission rarely reaches $F=1$
beyond $\simeq 2000\rm\,km\,s^{-1}$.  On average the maximum
correction is around 10 per cent, but it can be up to a 25 per cent
for some lines-of-sight at the edge of the proximity zone.  The
correction shifts the CPDF to lower Doppler widths, effectively
narrowing the absorption lines as the continuum is lowered.  For
reference, the median Doppler parameters are $b_{\rm
  med}=30.0\rm\,km\,s^{-1}$ (with correction) and $31.6\rm\,
km\,s^{-1}$ (no correction).  Consequently, if the continuum on the
{\it observed} data has been placed high (the correction always
lowers the continuum) relative to the corrected synthetic data, we
will infer gas temperatures which are $\sim 2000\rm\,K$ too high.
This is clearly an important systematic to consider when the
statistical error bars approach the level of a few thousand degrees.

\subsubsection{Metals}

Lastly, we consider the impact of mis-identified metal lines on the
Doppler parameter CPDF.  A full treatment would require modelling the
distribution and abundance of metals in the IGM at $z=6$, which is
beyond the scope of this work.  However, assuming metal lines are
typically narrower than the \Lya lines, we may estimate the effect of
any metal contamination as follows.  In our spectrum of
SDSS~J0818$+$1722, there are roughly $\sim 2$ metal absorption lines
per $3000\rm\, km\,s^{-1}$ interval redward of the \Lya emission line.
Over the same interval in the proximity zone, by comparison, we
include 25 lines in our Doppler parameter CPDF.  We may therefore
estimate the effect of possible metal contamination by excluding the
two narrowest lines identified in the proximity zone of
SDSS~J0818$+$1722 from our analysis, $b=8.6\rm\,km\,s^{-1}$ and
$b=10.0\rm\,km\,s^{-1}$ ({\it cf.} the narrowest line in model C,
$b=7.9\rm\,km\,s^{-1}$).  This changes the median Doppler parameter
for SDSS~J0818$+$1722 from $b_{\rm med}=31.0\rm\,km\,s^{-1}$ to
$b_{\rm med}=31.3\rm\,km\,s^{-1}$.  We estimate that metal
contamination may then at most systematically {\it decrease} any
constraint on the IGM temperature in the proximity zone by $\sim
2000\rm\,K$.  As we shall see, the systematic uncertainties discussed
in this section are small compared to the statistical uncertainty for
an individual line-of-sight.  Consequently, we do not consider these
in our final analysis.  However, we note that future analyses with
smaller statistical errors will require a more detailed treatment of
continuum placement and metal contamination.

%%%%%%%%%%%%%%%%%%%%%%%%%%%%%%%%%%%%%%%%%%%%%%%%%%%%%%%%%%%%%%%%%%%%%	
%%%%%%%%%%%%%%%%%%%%%%%%%% SECTION 5 %%%%%%%%%%%%%%%%%%%%%%%%%%%%%%%%
%%%%%%%%%%%%%%%%%%%%%%%%%%%%%%%%%%%%%%%%%%%%%%%%%%%%%%%%%%%%%%%%%%%%%

\section{Results}
\subsection{Test of the analysis procedure}
 
We are now ready to progress to the main result of this paper.
However, before proceeding further we briefly test our methodology.
We apply the analysis procedure described in
section~\ref{sec:analysis} to two synthetic lines-of-sight for which
we already know the ``true'' temperature.  In this way we may check if
our procedure correctly recovers the temperature of the IGM around a
quasar.

The test of our methodology is displayed in Figure~\ref{fig:Ttest}.
Each panel shows the constraint on $T_{0}$ for a synthetic
line-of-sight in the form of a cumulative distribution for $T_{0}$,
$P(<T_{0}|D_{\rm obs})$, computed using Eq.~(\ref{eq:Tdist}).  The
lines-of-sight are drawn from the B and F models which have
$T_{0}^{\rm true}=17\,500\,\rm K$ and $T_{0}^{\rm true}=34\,400\rm\,
K$, respectively.  These temperatures are represented by the filled
circles in Figure~\ref{fig:Ttest}.  The shaded range in both panels
display the 68 per cent confidence intervals around the inferred
median value of $T_{0}$.  Due to the limited statistical power of a
single line-of-sight the uncertainties on the constraints are large,
$T_{0}=18\,900\pm^{4400}_{6900}\rm\,K$ ($\pm^{12700}_{5000}\rm\,K$)
and $32\,400 \pm ^{4100}_{6000}\rm\,K$ ($\pm^{5600}_{10100}\rm\,K$) at
68 (95) per cent confidence around the median.  Nevertheless, it is
encouraging that our procedure correctly distinguishes between models
with different temperatures.

\subsection{Measurement of $T_{0}$ for SDSS~J0818$+$1722}
  
\begin{figure}
\begin{center}
  \includegraphics[width=0.45\textwidth]{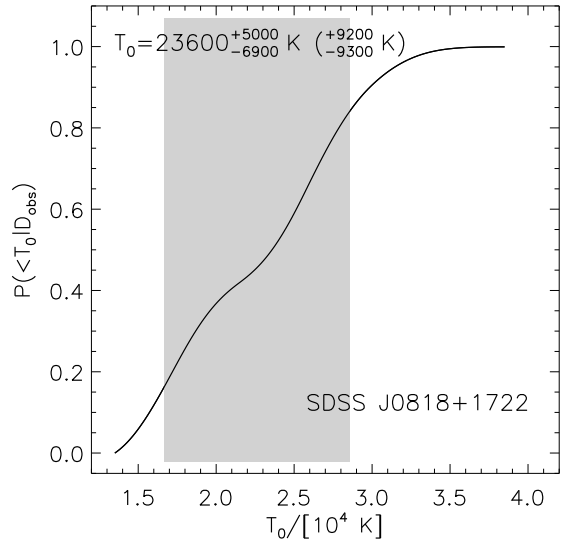}
\vspace{-0.3cm}
\caption{Constraint on the temperature of the IGM at mean density at
  $z=6$ in the proximity zone of the quasar SDSS J0818+1722.  The solid
  curves show the cumulative distribution for $T_{0}$,
  $P(<T_{0}|D_{\rm obs})$, obtained using Eq.~(\ref{eq:Tdist}), while
  the shaded region indicates the 68 per cent confidence interval
  around the median.  The numerical value in the top left of the panel
  corresponds to the recovered median temperature at mean density and
  the 68 (95) per cent confidence intervals.}
\label{fig:Tsdss}
\end{center}
\end{figure}

Figure~\ref{fig:Tsdss} displays the cumulative distribution for
$T_{0}$, $P(<T_{0}|D_{\rm obs})$, derived for the line-of-sight
SDSS~J0818$+$1722 at $z=6$.  We find a median temperature of
$T_{0}=23\,600\pm^{5000}_{6900}\rm\,K$ ($\pm^{9200}_{9300}\rm\,K$),
which lies in the middle of our simulation grid.  The CPDF flattens
towards high temperatures, indicating that temperatures of $T_{0}\ga
33\,000\rm\,K$ are inconsistent with the observed Doppler parameter
data.  The lower limit on $T_{0}$ is influenced by our choice of prior
for $T_{0}$ following \HI and \HeII reionisation, $13\,500 \rm\,K \leq
T_{0} \leq 38\,500\rm\,K$ ({\it e.g.}  \citealt{McQuinn09}).  However,
the gradient of the CPDF does begin to flatten towards the lower edge
of the simulation grid, suggesting the probability of temperatures
below $13\,500\rm\,K$ is fairly low. 

\subsection{Consistency check using proximity zone sizes}

In an analysis of observed proximity\footnote{WBH08 refer to these
  regions as highly ionised ``near-zones'', reflecting the fact that
  it can be difficult to determine whether these transmission windows
  are due to \HII regions embedded in a significantly neutral IGM or a
  region of enhanced ionisation ({\it i.e.}  the proximity effect) in
  an otherwise highly ionised IGM at $z>6$
  (\citealt{BoltonHaehnelt07,Maselli07,Lidz07}).  However, in this
  work we mainly concentrate on the ionised IGM at $z \leq 6$, and so
  for simplicity we refer to these regions as proximity zones
  throughout.}  zone sizes, WBH08 found the {\it absolute} sizes of
the proximity zones in their RT models were systematically smaller
than the observational data.  Based on their radiative transfer
modelling of the proximity zones WBH08 further found that the size of
the proximity zones should depend primarily on the background \HI
photo-ionisation rate and the temperature of the IGM, where the latter
impacts on the ionisation state of hydrogen through the \HII
recombination coefficient, $\alpha_{\rm HII} \propto T^{-0.7}$.  WBH08
therefore suggested that the IGM temperature in quasar proximity zones
at $z\simeq 6$ may be as high as $T_{0} \simeq 40\,000\rm\,K$, perhaps
resulting from the recent reionisation of \HI and \HeII by a hard
ionising spectrum.  Imposing a high temperature of $T_{0} \simeq
40\,000\rm\,K$ was found to resolve the discrepancy between the sizes
of observed and simulated proximity zones. Our results, albeit for a
single quasar proximity zone, are inconsistent with this
interpretation.  If our constraint is representative of the other
$z\simeq 6$ lines-of-sight, an alternative explanation would be
required to explain the proximity zone sizes.

\begin{figure}
\begin{center}
  \includegraphics[width=0.43\textwidth]{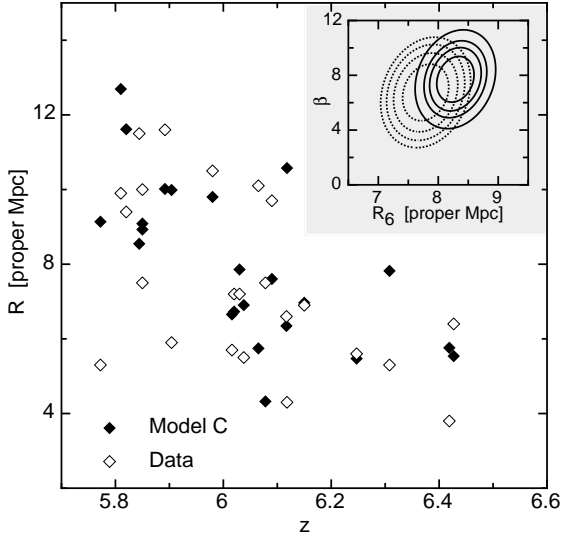}
\vspace{-0.2cm}
\caption{Proximity zone sizes (proper Mpc) measured from synthetic
  spectra (filled diamonds) compared to observational data taken from
  \citet{Carilli10}.  The simulated spectra are constructed from model
  C, which has gas temperatures consistent with our constraint from
  the proximity zone of SDSS~J0818$+$1722.  Following \citet{Fan06b}
  and \citet{Carilli10}, the sizes have been rescaled to a common AB
  magnitude of $M_{1450}=-27$ by assuming the proximity zone size is
  proportional to ${\dot N}^{1/3}$.  The inset displays the best fit
  linear relation to the observed and synthetic data, parameterised as
  $R=R_{6}-\beta(z-6)$.  The dotted and solid contours in the
  represent the 68, 84, 91 and 97 per cent bounds on single parameters
  for the observed and simulated data, respectively. }
\label{fig:sizes}
\end{center}
\end{figure}

In this paper we revisit the question of proximity zone size using our
new suite of high resolution simulations.  This analysis provides a
consistency check of our temperature constraints.  We have repeated
the proximity zone size analysis performed by WBH08 (see their section
5 for details) using our model C simulations.  This model has
$T_{0}=21\,500\rm\,K$, similar to the median value derived for
SDSS~J0818$+$1722. The main difference between our work and WBH08 is
the much higher resolution of our hydrodynamical simulations.  WBH08
used a simulation with gas particle masses of $4.3\times 10^{7}\rm
\,h^{-1}M_{\odot}$ and a box size of $60\,h^{-1}\rm Mpc$. This rather
large volume and low resolution was chosen so that the largest halo in
the simulation volume was reasonably massive, with a mass of
$2.7\times 10^{12}h^{-1}M_{\odot}$.  Here we use much higher
resolution hydrodynamical simulations which resolve the low density
IGM (\citealt{BoltonBecker09}).  In this work we have demonstrated
that mass resolution is more important than the halo mass for
correctly resolving transmission from the \Lya forest in the quasar
proximity zone.  We also now use the observational data compiled by
\cite{Carilli10}, which include seven additional quasars and an
improved determination of systemic redshifts.

The results are displayed in Figure~\ref{fig:sizes}.  The filled
diamonds correspond to the simulation data, while the open diamonds
show the observational data compiled in \cite{Carilli10}.  Following
\cite{Fan06b} and \cite{Carilli10}, the sizes\footnote{Following the
  observational definition of \cite{Fan06b}, WBH08 define the extent
  of the proximity zones as the distance from the quasar at which the
  transmission falls below 10 per cent after smoothing by a top hat
  filter of width 20\AA.} have been rescaled to a common AB magnitude
of $M_{1450}=-27$ by assuming the proximity zone size is proportional
to ${\dot N}^{1/3}$, where $\dot N$ is defined by Eq.~(\ref{eq:Ndot}).
Since our hydrodynamical simulation outputs are all at $z=6.01$, we
must also rescale the physical gas density in the simulations by
$(1+z)^{3}$ to match the redshifts of the quasars, $5.77 \leq z \leq
6.43$.  The inset shows the constraints on the amplitude and slope of
the proximity zone evolution using the parameterisation
$R=R_{6}-\beta(z-6)$, where $R_{6}$ is the value of the proximity zone
size at $z=6$ and $\beta$ is the slope of the evolution.  The dotted
and solid contours represent the 68, 84, 91 and 97 per cent bounds on
single parameters for the observed and simulated data, respectively.
We find that the simulated data are in reasonable agreement with the
observational data without having to arbitrarily impose high gas
temperatures as in WBH08.  This is because the much higher mass
resolution of our hydrodynamical simulation increases the transmission
from the underdense regions. These dominate the transmission towards
the edge of the proximity zone, increasing the extent of these
regions.  We have also verified that $T_{0}>33\,000\rm\,K$ produces
proximity zone sizes which are too large by performing the same
analysis using model G.  Our temperature constraint from
SDSS~J0818$+$1722 is therefore consistent with the rapid size
evolution of proximity zones at $z\simeq 6$, providing a useful check
of our modelling and results.  As already noted by WBH08, the rapid
evolution in the sizes of observed proximity zones also suggests that
the IGM in the environment of bright quasars is already highly ionised
by $z\simeq 6$.

%%%%%%%%%%%%%%%%%%%%%%%%%%%%%%%%%%%%%%%%%%%%%%%%%%%%%%%%%%%%%%%%%%%%%	
%%%%%%%%%%%%%%%%%%%%%%%%%% SECTION 6 %%%%%%%%%%%%%%%%%%%%%%%%%%%%%%%%
%%%%%%%%%%%%%%%%%%%%%%%%%%%%%%%%%%%%%%%%%%%%%%%%%%%%%%%%%%%%%%%%%%%%%

\section{Implications for the reionisation history around SDSS~J0818$+$1722}

\begin{figure*}
\centering
\begin{minipage}{180mm}
\begin{center}
\psfig{figure=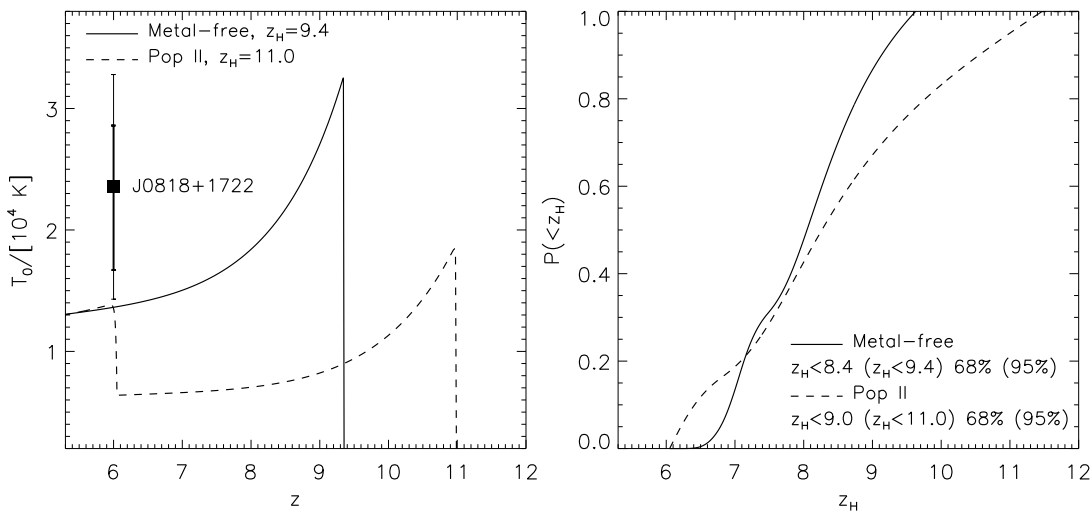,width=0.85\textwidth}

\vspace{-0.3cm}
\caption{{\it Left:} The IGM temperature at mean density, $T_{0}$, in
  a localised patch of the IGM as a function of redshift for two
  different instantaneous reionisation models.  The solid curve
  displays the temperature evolution following \HI and \HeII
  reionisation at $z_{\rm H}=z_{\rm He}=9.4$ by a very hard spectrum,
  representative of massive metal-free stars.  The dashed curve
  assumes $z_{\rm H}=11.0$ and corresponds to a much softer spectrum
  representative of population II sources.  Note that \HeII
  reionisation in this model is caused by the quasar itself at $z_{\rm
    He}=6.05$ ({\it i.e.} $t_{\rm Q}=10^{7}\rm\,yr$ before $z=6$).
  The filled square shows our constraint on the temperature in the
  proximity zone of SDSS~J0818$+$1722.  The thick (thin) error bars
  represent the 68 (95) per cent confidence interval around the
  median. {\it Right:} The cumulative probability distribution for the
  redshift of \HI reionisation around SDSS~J0818$+$1722 we derive from
  the two different models.  Our temperature measurement is consistent
  with \HI reionisation occurring at $z_{\rm H}<9.0$ ($z_{\rm
    H}<11.0$) at 68 (95) per cent confidence for the population II
  model, and $z_{\rm H}<8.4$ ($z_{\rm H}<9.4$) for the metal-free
  spectrum.  Note, however, that harder ionising spectra would
  increase both of these upper limits.}
\label{fig:tevol}
\end{center}
\end{minipage}
\end{figure*}

This work has focused on measuring the IGM temperature within close
proximity to a quasar.  Although this means we are unable to place
constraints on the thermal state of the ``average'' IGM as previous
studies have done at lower redshift ({\it e.g.}
\citealt{Schaye00,Ricotti00,McDonald01,Zaldarriaga02,Lidz09}), it does
enable us to study the IGM thermal history in the quasar's immediate
vicinity (\citealt{MiraldaRees94,Theuns02c,Meiksin09b}).  The
proximity zone temperature will encode information on when the IGM
near SDSS~J0818$+$1722 was first reionised.  The long timescale
associated with adiabatic cooling in the low density IGM enables the
gas to retain a ``fossil record'' of its initial temperature following
reionisation (\citealt{HuiGnedin97,Theuns02,HuiHaiman03}).  Adopting
an approach similar to that employed by \cite{HuiHaiman03}, we may
therefore use our temperature measurement to place an {\it upper
  limit} on the possible redshift of \HI reionisation around
SDSS~J0818$+$1722.  We pose the following question: ``What is the
maximum amount of time available for the IGM to cool following \HI
reionisation before the temperature becomes inconsistent with our
$T_{0}$ constraint at $z=6$?''

To address this question we shall compute the thermal evolution of a
gas parcel at mean density in the presence of a power law ionising
spectrum ({\it e.g.}  \citealt{HuiHaiman03,FurlanettoOh09}).  Our code
follows photo-ionisation and heating, collisional ionisation,
radiative cooling, Compton cooling and adiabatic cooling for six
species (H~$\rm \scriptstyle I$, H~$\rm \scriptstyle II$, He~$\rm
\scriptstyle I$, He~$\rm \scriptstyle II$, He~$\rm \scriptstyle III$,
$\rm e^{-}$).  We use the rates compiled in \cite{BoltonHaehnelt07}
with the exception of the case B recombination and cooling rates of
\cite{HuiGnedin97} and the photo-ionisation cross-sections of
\cite{Verner96}.

We consider two possibilities for the spectra of \HI ionising sources
(and hence IGM temperatures following \HI reionisation) which should bracket
the plausible range of possibilities.  Following \cite{HuiHaiman03},
we base the first model on the very massive, metal-free stellar
spectrum presented by \cite{Bromm01}: $J_{\nu}\propto \nu$ for
$\nu_{\rm HI} \leq \nu \leq \nu_{\rm HeI}$, $J_{\nu}\propto \nu^{0}$
for $\nu_{\rm HeI} \leq \nu \leq \nu_{\rm HeII}$ and $J_{\nu}\propto
\nu^{-4.5}$ for $\nu\geq \nu_{\rm HeII}$, where $\nu_{\rm i}$
corresponds to the ionisation threshold for species $i$.  This
spectrum is very hard, and will result in significant \HI {\it and}
\HeII photo-heating during reionisation.  Note that since this model
reionises \HeII at high redshift, it is inconsistent with \HeII
reionisation completing at lower redshift unless the spectrum softens
significantly and \HeIII partially recombines again before $z\simeq
3$.  The second model, a more conventional population II spectrum, is
based on the model of \cite{Leitherer99} for a galaxy of age 500 Myr
with a continuous star formation rate, a Salpeter IMF and metallicity
$Z=0.2Z_{\odot}$.  In this instance $J_{\nu} \propto \nu^{-3}$ below
the \HI ionisation threshold.  This soft spectrum does not
significantly reionise He~$\rm \scriptstyle II$.  Note that if the IGM
is instead reionised by an earlier round of quasar activity, or if
quasars are responsible for primarily driving \HI reionisation
(\citealt{VolonteriGnedin09}, but see
\citealt{Srbinovsky07,Willott09}), the ionising spectrum and resulting
IGM thermal history will be intermediate between these two cases.

To account for filtering of the ionising radiation as it propagates
through optically thick gas, we assume the mean free path for
ionising photons in a clumpy IGM has a frequency dependence
$\lambda_{\nu}\propto \nu^{1.5}$
(\citealt{ZuoPhinney93,MiraldaEscude03}).  This hardens the
intrinsic spectra by $\alpha_{\rm s} \rightarrow \alpha_{\rm s} -
1.5$, and is intermediate between the optically thin case and the
spectral modification expected for a uniform, optically thick IGM,
$\alpha_{\rm s} \rightarrow \alpha_{\rm s} - 3$
(\citealt{AbelHaehnelt99}).  Lastly, we normalise both models to give
an \HI photo-ionisation rate of $\Gamma_{\rm HI}=1.9\times
10^{-13}\rm\,s^{-1}$, which is consistent with the upper limit inferred
from the observed \Lya forest opacity at $z=6$
(\citealt{BoltonHaehnelt07b}).

Two example thermal histories are displayed in the left hand panel of
Figure~\ref{fig:tevol}, along with our $T_{0}$ constraint for the
proximity zone of SDSS~J0818$+$1722.  The thick (thin) error bars
correspond to the 68 (95) per cent confidence intervals around the
median $T_{0}$. We assume instantaneous \HI and \HeII reionisation for
the metal-free spectrum at $z_{\rm H}=z_{\rm He}=9.4$ (solid curve).
The population II model (dashed curve) instead has $z_{\rm H}=11.0$.
To account for \HeII photo-heating by the hard spectrum of the quasar
itself we also include \HeII reionisation at $z_{\rm He}=6.05$ ({\it
  i.e.} $t_{\rm Q}=10^{7}\rm\,yr$ before $z=6$) for a quasar like
spectrum with $J_{\nu}\propto \nu^{-1.5}$ (hardened to $J_{\nu}\propto
\nu^{0}$).  This leads to a temperature boost during \HeII
reionisation of $\sim 8000\rm\,K$, consistent with the results of our
radiative transfer models. 

We stress that instantaneous reionisation is an unphysical model for
the ionisation and thermal history of the entire IGM during
inhomogeneous \HI reionisation.  However, our temperature constraint
is for the volume within $33$ Mpc of a bright quasar.  If photoheating
is the dominant heating process during the epoch of reionisation, we
should nevertheless obtain an upper limit on the redshift when the
bulk of the reionisation occurred around the quasar in this
approximation. Numerical simulations indicate such biased regions are
amongst the first patches of the IGM to be reionised ({\it e.g.}
\citealt{Iliev06b,Shin08,Finlator09}).  A constraint on $z_{\rm H}$
within 33 Mpc SDSS~J0818$+$1722 might then also give a constraint for
the {\it global} onset of reionisation.  However, a joint analysis
with other observational constraints including the \Lya forest opacity
(\citealt{Fan06b}) and the electron scattering optical depth
(\citealt{Dunkley09}) will be required for a consistent interpretation
of the temperature data (\citealt{ChoudhuryFerrara06,Pritchard09}).
In this work, {\it our reionisation constraints apply to the IGM
  around SDSS~J0818$+$1722 only.}

The right hand panel of Figure~\ref{fig:tevol} displays the cumulative
probability distribution for the redshift of \HI reionisation,
$P(<z_{\rm H})$, for these two models using our $z=6$ temperature
measurement.  Assuming \HeII reionisation by the quasar boosts the IGM
temperature by $\sim 8000\rm\,K$, we find the population II model is
consistent with \HI reionisation occurring at $z_{\rm H}<9.0$ ($z_{\rm
  H}<11.0$) at 68 (95) per cent confidence.  However, a harder quasar
spectrum ($\alpha_{\rm s}<1.5$), and hence a larger temperature boost
during \HeII reionisation, would weaken this constraint considerably.
On the other hand, if \HI and \HeII reionisation around the quasar are
driven primarily by a generation of metal free stars with hard
spectra, we find $z_{\rm H}<8.4$ ($z_{\rm H}<9.4$) at 68 (95) per cent
confidence.

We caution that the astrophysical uncertainties in the modelling
presented here are significant; the exact spectrum of the ionising
sources are rather uncertain at $z>6$.  Harder ionising spectra would
increase these upper limits.  This study nevertheless demonstrates
that even a single temperature measurement at $z=6$ can give an
interesting constraint on the redshift of \HI reionisation around
SDSS~J0818$+$1722.  Although computationally expensive, a larger set
of simulations will allow for a much more refined calibration of the
measurement.  Another obvious extension is to analyse additional
lines-of-sight ({\it e.g.}  \citealt{Becker05}).  Independent
measurements along separate lines-of-sight will aid in reducing the
statistical uncertainty on any averaged measurement.  The amount of
scatter in the temperature from one line-of-sight to the next may also
provide some insight into the topology and timing of \HI reionisation
globally.  Improving existing constraints on the IGM thermal history
at $z<4.5$, as well as adding to constraints at higher redshift, will
thus provide extremely valuable insight the epochs of both \HeII and
\HI reionisation.

%%%%%%%%%%%%%%%%%%%%%%%%%%%%%%%%%%%%%%%%%%%%%%%%%%%%%%%%%%%%%%%%%%%%%	
%%%%%%%%%%%%%%%%%%%%%%%%%% SECTION 7 %%%%%%%%%%%%%%%%%%%%%%%%%%%%%%%%
%%%%%%%%%%%%%%%%%%%%%%%%%%%%%%%%%%%%%%%%%%%%%%%%%%%%%%%%%%%%%%%%%%%%%

\section{Conclusions}

In this paper we present the first direct measurements of the IGM
temperature around a quasar at $z=6$.  We use a combination of high
resolution hydrodynamical simulations combined with a line-of-sight
radiative transfer implementation to model the thermal state of the
IGM in a quasar proximity zone.  Previous theoretical studies have
suggested that the IGM temperature close to a quasar will be sensitive
to the prior ionisation state of hydrogen and helium, as well as the
intrinsic quasar spectrum
(\citealt{MiraldaRees94,BoltonHaehnelt07,Lidz07}).  We demonstrate
here that the Doppler parameter CPDF obtained from the quasar
proximity zone provides a sensitive probe of the IGM temperature,
enabling us probe the thermal and ionisation history of the IGM within
the quasar's vicinity.

Our observational data set consists of a single high resolution, high
$S/N$ Keck/HIRES spectrum of the quasar SDSS~J0818$+$1722 at $z=6$.
We perform identical Voigt profile analyses on the observed and
synthetic data to obtain the Doppler parameter CPDF.  The simulations,
which use a range of self-consistent thermal histories, are used to
calibrate the line width measurements obtained from the observational
data.  After verifying our method for obtaining temperature
constraints, we proceed to obtain
$T_{0}=23\,600\pm^{5000}_{6900}\rm\,K$ $(\pm^{9200}_{9300}\rm\,K)$ at
68 (95) per cent confidence around the median for the proximity zone
of SDSS~J0818$+$1722.  Due to the small size of the data set, the
error bars are dominated by statistical uncertainty; we have verified
that the most important systematic uncertainties remain small in
comparison.  We find our simulations are also in good agreement with
the rapid evolution observed in proximity zone sizes at $z\simeq 6$
when adopting temperatures consistent with our direct constraint.
This provides a consistency check of our results, and further suggests
that the IGM in the environment of bright quasars is highly ionised by $
z\simeq 6$ (\citealt{Wyithe08,Maselli09}).

Finally, we use our temperature constraint to explore the implications
for the IGM reionisation history around SDSS~J0818$+$1722 ({\it e.g.}
\citealt{Theuns02,HuiHaiman03}).  We consider two simple models for
the thermal history assuming that reionisation in the proximity zone
occurred instantaneously.  Assuming \HI reionisation around
SDSS~J0818$+$1722 is driven by ionising sources with soft spectra,
typical of population II stars, we infer $z_{\rm H}<9.0$ $(11.0)$ at
68 (95) per cent confidence if photo-heating is the dominant heating
mechanism during the epoch of reionisation.  However, if the IGM is
instead reionised by a population of massive metal-free stars,
characterised by very hard ionising spectra, we obtain a tighter limit
of $z_{\rm H}<8.4$ $(9.4)$.  An \HI reionisation epoch beginning at
higher redshifts produces temperatures which are too low with respect
to our constraint unless the \HI ionising sources or the quasar itself
have spectra significantly harder than typically assumed.

\section*{Acknowledgements}

We thank Benedetta Ciardi and Peng Oh for valuable conversations
during the course of this work.  The hydrodynamical simulations used
in this work were performed using the Darwin Supercomputer of the
University of Cambridge High Performance Computing Service
(http://www.hpc.cam.ac.uk/), provided by Dell Inc. using Strategic
Research Infrastructure Funding from the Higher Education Funding
Council for England.  JSB acknowledges the support of an ARC
Australian postdoctoral fellowship (DP0984947), and GDB thanks the
Kavli foundation for financial support.

\end{document}